%% file: main.tex
\documentclass[conference]{IEEEtran}
\IEEEoverridecommandlockouts
\usepackage{cite}
\usepackage{amsmath,amssymb,amsfonts}
\usepackage{algorithmic}
\usepackage{graphicx}
\usepackage{textcomp}
\usepackage{xcolor}
\usepackage{quantikz}
\usepackage{tikz}
\usepackage{comment}
\usepackage{svg}
\usepackage{caption}
\usepackage{subcaption}
\usepackage{csvsimple}
\usepackage{standalone}
\usepackage{subcaption}
\usepackage{subfiles}
\usepackage{color, soul}
\usepackage{float}
\usepackage{url}
\graphicspath{{\subfix{figs/}}}
\def\BibTeX{{\rm B\kern-.05em{\sc i\kern-.025em b}\kern-.08em
    T\kern-.1667em\lower.7ex\hbox{E}\kern-.125emX}}
\begin{document}

\title{Improving Qubit Routing by Using Entanglement Mediated Remote Gates
\thanks{This work was supported by Mitacs through the Mitacs Business Strategy Initiative program.
        We acknowledge the support of the Natural Sciences and Engineering Research Council of Canada (NSERC). Part of this work was supported by the NSERC‐Collaborative Research and Training Experience program QSciTech. We acknowledge the support of the Ministère de l'Enseignement Supérieur du Québec.}
}

\author{\IEEEauthorblockN{Gurleen Padda \IEEEauthorrefmark{1}, Edwin Tham\IEEEauthorrefmark{2}, Aharon Brodutch\IEEEauthorrefmark{2} and Dave Touchette\IEEEauthorrefmark{1}}\\
\IEEEauthorblockA{\IEEEauthorrefmark{1} \textit{Department of Computer Science and Institut Quantique, Université de Sherbrooke}, Sherbrooke, QC, Canada\\Email: \{gurleen.padda, dave.touchette\}@usherbrooke.ca}\\
                    
\IEEEauthorblockA{\IEEEauthorrefmark{2}\textit{IonQ Canada}, Toronto, ON, Canada\\Email: \{tham, brodutch\}@ionq.co}}

\maketitle
\thispagestyle{plain}
\pagestyle{plain}
\begin{abstract}
    Near-term quantum computers often have connectivity constraints, i.e. restrictions, on which pairs of qubits in the device can interact. Optimally mapping a quantum circuit to a hardware topology under these constraints is a difficult task. While numerous approaches have been proposed to optimize qubit routing, the resulting gate count and depth overheads of the compiled circuits remain high due to the short-range coupling of qubits in many near-term devices. Resource states, such as Bell or Einstein-Podolsky-Rosen (EPR) pairs, can be used to mediate operations that facilitate long-range interactions between qubits. In this work, we studied some of the practical trade-offs involved in using resource states for qubit routing. We developed a method that leverages an existing state-of-the-art compiler to optimize the routing of circuits with both standard gates and EPR mediated remote controlled-NOT gates. This was then used to compile different benchmark circuits for a square grid topology, where a fraction of the qubits are used to store EPR pairs. We demonstrate that EPR-mediated operations can substantially reduce the total number of gates and depths of compiled circuits when used with an appropriate optimizing compiler that accounts for practical overheads. Our results highlight the relevance of developing efficient compilation tools that can integrate EPR-mediated operations.
\end{abstract}

    \section{Introduction}
        Executing a quantum circuit on a Noisy Intermediate Scale Quantum (NISQ) device~\cite{preskill_2018} requires performing a number of pre-processing steps in order to render the circuit compatible with the device. As part of this pre-processing, it is necessary to ensure that the qubit connectivity constraints of the device are respected by the circuit. This may require adding additional gates to the circuit to exchange or move information between qubits. This is known as qubit routing~\cite{siraichi_2018}. 
    
        Given the low fidelities of gates and short coherence times of qubits in NISQ devices~\cite{preskill_2018, lau_2022}, the number of gates and the depth of circuits run on them should be minimized. While various approaches have been proposed to optimize qubit routing~\cite{sabre, cnot_synth, tket, siraichi_2018, zulehner2018, pyzx, maslov, gheorghiu}, the resulting gate count and depth overheads of the compiled circuits remain high due to the short range coupling of qubits in these devices~\cite{leymann2020, herbert2020}. Resource states, such as Bell or Einstein-Podolsky-Rosen (EPR) pairs, can be used to mediate operations, such as quantum teleportation~\cite{teleportation} and remote gates~\cite{collins, nielson_r_cx, sorensen}, that facilitate the long-range communication of quantum information. Access to these resource states might therefore be  advantageous for qubit routing, as they can augment the effective connectivity of the underlying hardware. There is a need to better understand the potential improvement this can deliver when routing practical quantum circuits.

        An approach that uses teleportation for qubit routing, and can achieve a depth advantage over swap based routing~\cite{alon1993routing} was proposed by Devulapalli et al.~\cite{devulapalli2022}. However, the number of gates that it uses scales with the length of the path along which the long-range teleportation is performed, and this would therefore not yield an advantage in gate count over swap based routing. In addition, the protocol requires that there be an ancilla qubit available for every data qubit of the circuit, resulting in an $O(n)$ ancilla qubit overhead, where $n$ is the number of qubits in the circuit. Another approach was proposed by Hillmich et al.~\cite{hillmich2021}, in which any pairs of qubits available on the device that are not needed as data qubits are used to form an EPR pair, which is then used with teleportation for qubit routing. While this results in an advantageous ancilla qubit overhead, the distribution of these resource states throughout the device is not optimized, and it is unlikely that this would outperform the best methods currently available for qubit routing. In both of these works, qubits that form an EPR pair must at one point have been neighbours in the device so that they could become entangled by quantum gates. 

        Technology that can distribute resource states, such as EPR pairs, across a network of quantum processors, has been developed and demonstrated in recent years~\cite{stephenson, monroe} (see also Sec.~\ref{sec:gen} below).
        These EPR pair generation schemes, while practically challenging, continue to be improved nevertheless, as they are widely considered to be a necessary component in the scaling of quantum computers. Given that developing this technology is a significant challenge, it is important that all of it's potential advantages be well understood.
        The goal of this work is to demonstrate the possible benefits and trade-offs of using EPR states as a means to improve qubit routing. We envision distributing EPR pairs throughout distant sites within a single quantum processor, and investigate the potential advantages of using EPR mediated remote gates~\cite{remote_cx} for qubit routing. 
        
        While it is evident that this could help improve the connectivity of the qubits in the device, using this increased connectivity incurs many additional costs.
        It is \emph{not} always obvious that the increased connectivity will consistently result in a net benefit for qubit routing.
        Notably, these EPR pairs occupy physical qubits in the device that could have otherwise been used to execute the circuit, and can reduce the effective connectivity between qubits in the neighborhood of an EPR site. Furthermore, using an EPR pair to perform a remote gate requires inserting additional gates into the circuit, and a net reduction in the gate counts and depths of the final compiled circuits from their use is not a priori obvious. 
        In this work, we demonstrate that they can in fact offer a substantial improvement to these costs.
        
        Section~\ref{sec:background} presents the necessary background on qubit routing, circuit compilation, EPR-mediated remote CX gates, and generating EPR pairs. In Section~\ref{sec:techniques}, we describe a method that leverages an existing state-of-the-art, heuristic based compiler, to optimize the use of both standard gates and EPR mediated remote CX gates to route circuits. This method was then used to compile different benchmark circuits for a device whose native qubit connectivity is that of a square grid. Ancilla qubits, that were used to store EPR pairs, were placed at regular intervals along the edges of the grid, resulting in a $O(\sqrt{n})$ ancilla qubit overhead. The benchmark results are given in Section~\ref{sec:results}. Contingent on the compiler’s ability to perform fidelity-weighted (noise adaptive) routing, we show that using remote CX gates can significantly reduce both the final standard CX gate counts and depths of the compiled circuits, and that this advantage scales with the size of the architecture. Our results underscore the potential value in creating efficient compilation tools capable of incorporating EPR mediated operations. Similar techniques can be adapted to route qubits between compute units that are sparsely connected~\cite{tham2022}; a scenario that appears increasingly likely as quantum computers scale, and modularity becomes paramount.
        
    \section{Background} \label{sec:background}
        \subsection{Qubit Routing and Optimizing Compilers}
            An optimizing compiler takes a circuit and a description of a device as input, and returns a compiled circuit that is logically equivalent to the input circuit, and is compatible with the device. It also attempts to optimize how the compiled circuit will perform when executed on the device, which usually includes minimizing its gate count and depth. Designing such an optimizing compiler is a difficult task~\cite{botea_2018, ito2023}, and there are many efforts underway to improve existing techniques and design new ones. 
            
            Given a quantum circuit, all gates that involve multiple qubits can be decomposed into a combination of single-qubit and CX gates~\cite{nielsen_chuang_2010}. We will assume from here on, when performing qubit routing on a circuit, that it has already been decomposed into single qubit and CX gates. 

            \subsubsection{Coupling graphs}
                A common way to represent the connectivity of the physical qubits in a quantum device is through the use of a coupling graph, where the nodes of the graph represent the qubits, and the edges connect qubits that can perform two qubit gates. The fidelity of the gates performed between different pairs of qubits in the device may vary, and this can be represented in the coupling graph by assigning weights to the edges. For near-term quantum computers, particularly for superconducting architectures, physical qubit connnectivity remains low and the coupling graphs are often planar~\cite{chow2021ibm, arute2019quantum}.

            \subsubsection{The qubit routing problem}
                Typically, quantum circuits are designed without a particular device topology in mind, relying on the assumption that any pair of qubits can execute a two qubit gate. However, this assumption does not hold for near term architectures. Qubit routing is a step in the compilation process that allows a quantum circuit, initially designed on a set of virtual qubits, to be executed on a physical device such that the qubit connectivity constraints of the device are respected. The process of qubit routing first involves mapping the virtual qubits of a quantum circuit to the physical qubits of the target quantum device. An optimal initial layout should minimize the number of gates that will then have to be added to the circuit to connect the qubits involved in two qubit gates that, according to the chosen layout, cannot be performed due to the physical qubit connectivity constraints. Once the initial mapping is done, the qubit routing process then applies routing techniques to establish viable paths between these qubits, and inserts gates along these paths to resolve the lack of connectivity. The goal is to accommodate the physical qubit layout and connectivity constraints, while minimizing the gate count and depth overheads of the compiled circuit, in order to yield the best possible fidelity of quantum computations on the specific quantum device. 
            \subsubsection{Optimizing qubit routing}   
                Qubit routing has been shown to be an NP-hard problem~\cite{ito2023}, and various techniques have been developed to minimize the gate count and depth overheads of routed circuits~\cite{sabre, cnot_synth, tket, siraichi_2018, zulehner2018, pyzx, maslov, gheorghiu}. A common technique is known as swap based routing~\cite{alon1993routing}. With this technique, swap gates, that can exchange the quantum state of two qubits, are inserted into the circuit. By applying swap gates between qubits located along paths in the coupling graph that connect distant qubits, these distant qubits can be moved closer together and become connected, which aligns the circuit with the connectivity of the qubits in the physical device. This technique is usually optimized using heuristic methods that take into consideration the shortest paths along which to insert these swaps, as well as how these swaps will affect future operations. Swapping two qubits can have contrasting effects: it may bring one of the qubits closer to some other qubit that it needs to perform a gate with, while simultaneously moving the other away from one it needs to interact with later on. Non-optimal placement of these swap gates may introduce many more instances of non-neighbouring qubits that need to perform two qubit gates, which would then require more swap gates to resolve. Each additional swap gate increases the overall depth and gate count of the circuit, and this can lead to significant overheads that scale with the size of the circuits, making them more susceptible to noise. 
                
                An alternative to swap based routing is topologically constrained synthesis~\cite{cnot_synth}. This approach re-synthesizes the circuit with the topology of the quantum device in mind. A quantum circuit (assuming non-unitary elements have been removed) is a decomposition of a unitary operation into elementary gates. Instead of using swap gates to resolve connectivity issues, topologically constrained synthesis focuses on finding an alternative decomposition of the unitary operation, to obtain a sequence of elementary gates that respects the device's connectivity constraints. This allows for a more efficient utilization of the available qubit connectivity and can reduce the number of gates needed to route the circuit. This is the routing method implemented by SoftwareQ's Steiner tree mapper, available through their compiler toolkit \texttt{staq}~\cite{staq}.
        
                Of the compilers currently available, \texttt{staq} offers tools that are well suited for our application. It offers support for fidelity-weighted or noise adaptive routing, where it is able to account for different fidelities at different edges in the coupling graph. It uses a weighted graph routing algorithm to connect nodes in the graph using paths whose edge fidelities sum to the highest possible value. This feature has proven to be particularly useful for our application.
        
        \subsection{EPR-mediated remote CX Gates}
            \begin{figure}
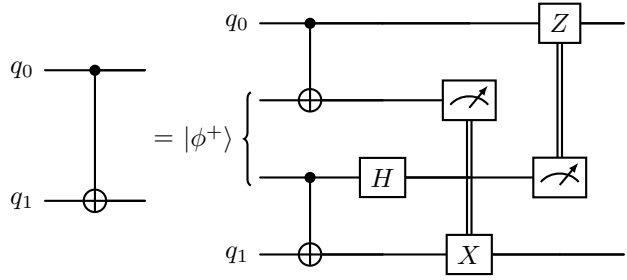

                \centering
                \includestandalone[scale=0.8]{remote_cx}
                \caption{%
                    A circuit that performs a remote CX gate between two data qubits using an EPR pair. Once the remote CX operation is performed, the state of qubits $q_{0}$ and $q_{1}$ is the same as that that would have resulted from performing a standard CX gate between them.
                }%
                \label{fig:remote_cx_circuit}
            \end{figure}

            Detailed presentations of the EPR-mediated remote CX gate can be found in~\cite{remote_cx,tham2022}. Similar to quantum teleportation, this operation uses an EPR pair to perform a CX gate between distant qubits. A decomposition of the remote CX gate is shown in Figure~\ref{fig:remote_cx_circuit}. In the decomposition, a standard CX gate is performed between each data qubit and the half of the EPR pair it neighbours. The qubits that form the EPR pair are measured, and the measurement outcomes are used to determine whether corrective rotations should be applied. This operation integrates well with existing techniques and compilers, as it is simply a long range version of a gate that is already used during qubit routing. Assuming an EPR pair is already available, the depth and number of gates required to perform a remote CX gate is constant, and therefore independent of the distance across which the remote gate is performed. It does, however, have an increased cost in both depth and gate count as compared to a single standard CX gate. The compiler must be able to consider this trade-off and employ the remote CX gate only when it proves to be more advantageous than utilizing standard gates alone.
            
    \subsection{Generating EPR pairs}\label{sec:gen}
        In~\cite{devulapalli2022, hillmich2021}, entanglement was assumed to be generated at the logical level using additional gates and qubits. Here we assume that entangled resource states can be generated in a different, more direct process using photonic interconnects~\cite{Awschalom_2021, monroe}. Photonic interconnects are necessary components for transporting quantum information in a modular quantum computer. They are widely viewed as an essential technology for scaling quantum computers~\cite{Awschalom_2021}. The general method of operation for a photonic interconnect is to entangle qubits with photons, and then use a Bell type measurement to swap the entanglement from a pair of qubit-photon entangled states to a qubit-qubit EPR pair. Multiple physical mechanisms that can be used to perform both the initial qubit-photon entanglement, and the final measurement that heralds qubit-qubit entanglement, can be found in~\cite{monroe,Awschalom_2021,stephenson}.   

        At the moment, photonic interconnects have not yet been fully demonstrated within a fully operational quantum computer. There are three main issues that need to be resolved. 
        \begin{itemize}
            \item Impact on other qubits: There have been very few demonstrations where entanglement was generated on one qubit, while allowing a neighbouring qubit to remain available for storing quantum information~\cite{Hucul_2014}. This is a challenge due to the impact of the entangling operation on neighbouring qubits.
            \item Low rates/high latency: The rates at which qubit-photon entanglement can be generated are low for reasons that depend on the specific system used. In many cases, the method used for generating this entanglement is based on spontaneous emission, and the photon collection efficiency is very low (e.g.~\cite{stephenson}). In other cases, the system operates at microwave frequencies, and converting to optical frequencies is required. This is currently an extremely inefficient process~\cite{transduc}.
            \item Entanglement fidelity: Entanglement fidelity in a photonic interconnect is impacted by three main factors: the initial qubit-photon entanglement, the detection mechanism that projects the qubits into a joint EPR state, and the life-time of the qubit. The fidelity of the EPR pairs generated using the current state-of-the-art is far below that of two qubit gate fidelities and needs to be improved. 
        \end{itemize}
        
        We stress that this technology is necessary for scaling quantum computers~\cite{tham2022}, and that there is a need to understand where it can be used once it is mature enough.

    \section{Techniques} \label{sec:techniques}
        \subsection{Routing Model} \label{sec:model}
            We considered a device with the qubit connectivity of a square grid, where a fixed number of qubits on the device are reserved as ancilla qubits to form EPR pairs. In our model, fixed pairs of these qubits could become entangled through the use of an entanglement generating device. We imposed that these qubits could only form EPR pairs and be involved in remote CX operations. They could not interact otherwise with their neighbouring data qubits. This ensured that they could always be reset and entangled before the remote gate operation without erasing important information pertaining to the execution of the circuit. 
    
\begin{figure}
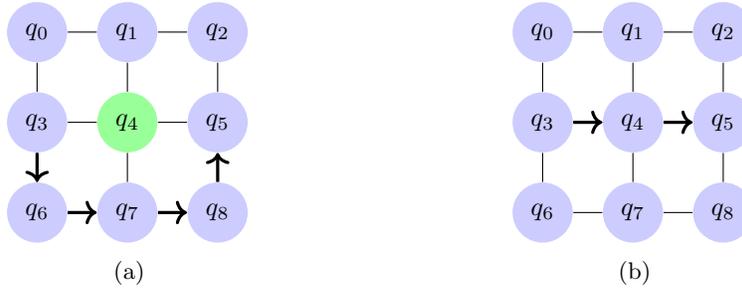

                \centering
                \begin{subfigure}{0.2\textwidth}{%
                        \hspace{0.55cm}
                        \includestandalone[scale=0.7]{tikz_hole}
                    }%
                    \caption{}
                    \label{fig:cm_hole}
                \end{subfigure}
                \begin{subfigure}{0.2\textwidth}{%
                \hspace{0.55cm}
                        \includestandalone[scale=0.7]{tikz_no_hole}
                    }%
                    \caption{}
                    \label{fig:cm_no_hole}
                \end{subfigure}
                \caption{%
                    \subref{fig:cm_hole} A path connecting qubits $q_{3}$ and $q_{5}$ would have to `go around' ancilla qubit $q_{4}$ in our model, resulting in a longer path. When it's not being used for a remote gate, qubit $q_{4}$ essentially becomes a `hole' in the graph, and reduces the effective local connectivity.
                    \subref{fig:cm_no_hole} With no ancilla qubit present, a shorter path connecting $q_{3}$ and $q_{5}$ can be found.
                }%
                \label{fig:cm_holes}
            \end{figure}

\begin{figure}
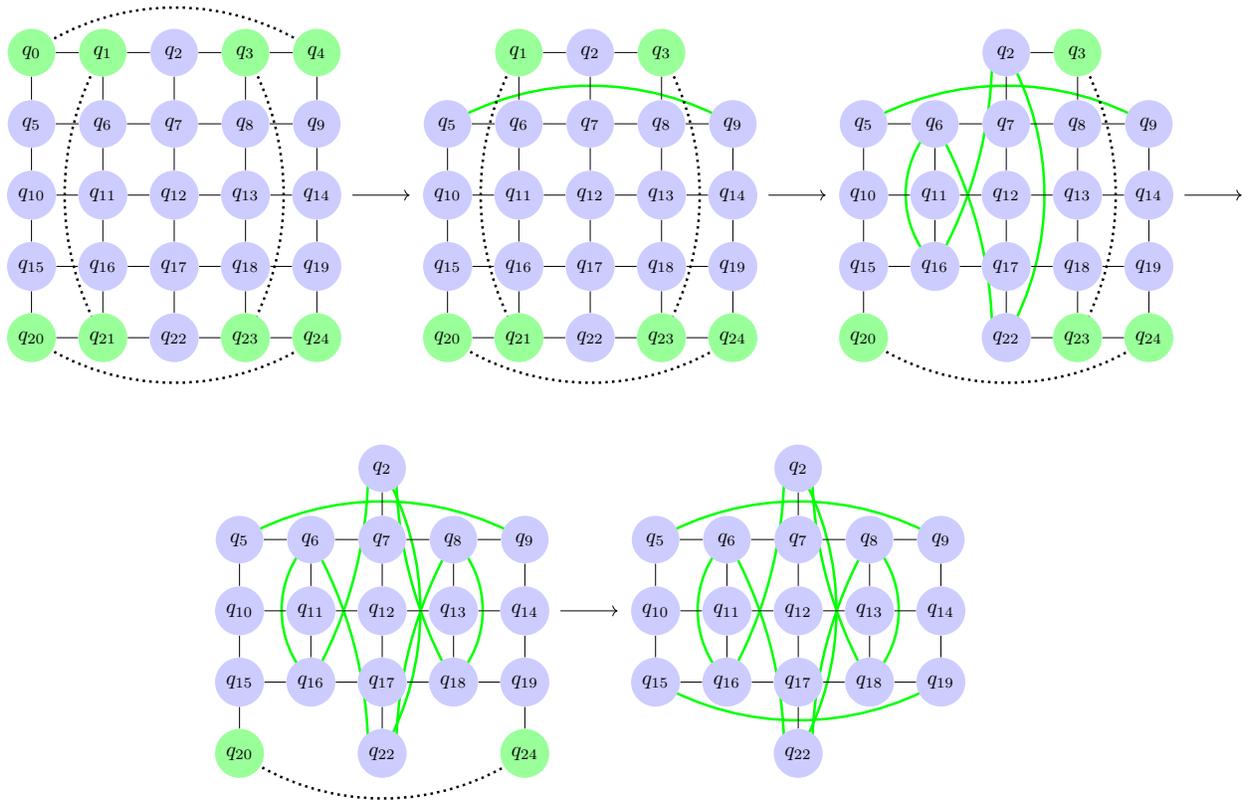

                \centering
                \includestandalone[width=0.5\textwidth]{tikz_epr}
                \caption{%
                    A coupling graph depicting the layout of ancilla qubits in our model for a $5 \times 5$ grid, and the augmented coupling graph generated from it. Black nodes represent ancilla qubits, and white nodes represent data qubits. The solid lines connect nodes that can perform two qubit gates, and the dotted lines connect qubits that are entangled. To generate the augmented graph on the right, the ancilla qubits that form EPR pairs are removed. For each EPR pair, the pairs of data qubits that they can perform remote CX gates with are connected by a solid edge. These augmented edges are the thicker edges shown in the figure on the right. Note that the augmented edge $q_{2}-q_{22}$ can result from EPR pairs across either $q_{1}-q_{21}$, or $q_{3}-q_{23}$.
                }%
                \label{fig:coupling_maps}
            \end{figure}

            While the EPR pairs formed by these ancilla qubits can help increase the long-range connectivity of the device, imposing that they can only be used to perform remote gates can reduce the effective local connectivity of the qubits, depending on where the ancilla qubits are placed. This is depicted in Figure~\ref{fig:cm_holes}. We chose to place the ancilla qubits at the edges of the grid, as this reduces the impact of this loss of local connectivity, while also obtaining maximal use of the long-range capability offered by remote gates. EPR pairs were then formed between ancilla qubits located at the ends of the same row or column. The chosen layout of ancilla qubits for a $5 \times 5$ grid is shown in Figure~\ref{fig:coupling_maps}. As the grids grow in size to fit larger circuits, additional EPR pairs are inserted along the edges at regular intervals. A mapping between the size of the circuit and size of the coupling graph used, as well as the number of EPR pairs distributed is given in Table~\ref{tab:grid_sizes}. Any data qubit neighbouring half of an EPR pair can perform a remote CX gate with any data qubit neighbouring the other half.
    
            \begin{table}[]
                \centering
                \begin{tabular}{|c|c|c|}\hline
                    \# qubits & Compiled \# qubits & \# EPR pairs\\\hline\hline
                    \begin{filecontents*}{data/grid_sizes.csv}
a,b,c,d
1-17,25,4,0.32
18-26,36,5,0.277777778
27-37,49,6,0.244897959
38-52,64,6,0.1875
53-67,81,7,0.172839506
68-84,100,8,0.16
85-105,121,8,
\end{filecontents*}
\csvreader[head to column names, late after line = \\]{data/grid_sizes.csv}{}{\a & \b & \c}
                    \hline
                \end{tabular}
                \caption{%
                    A mapping between the number of qubits in the benchmark circuit and the size of the coupling graph used, as well as the number of EPR pairs distributed.
                }%
                \label{tab:grid_sizes}
            \end{table}
        \subsection{Compilation Method} \label{sec:method}
            To investigate the effects of using EPR mediated remote gates for qubit routing, we opted to route practical quantum circuits under this model and assess the overheads incurred by the compiled circuits. To do this, it was necessary to have an optimizing compiler that could balance the trade-off between the increased cost of using a remote gate and the advantage of being able to perform the gate on distant qubits. The compilation method we describe here leverages an existing optimizing compiler that, as is, is not designed to compile circuits using both remote and standard gates. With this method, it was not necessary to build an optimizing compiler from scratch to explicitly target EPR-mediated architectures, though the results obtained using this approach could justify future work pursuing such a compiler project.
            
            In order to compile a circuit using this model, an augmented coupling graph was generated for the device. To generate this augmented graph, the nodes in the original coupling graph that represent ancilla qubits are removed. In their place, edges are added to the graph between qubits that can perform remote CX gates (see Figure~\ref{fig:coupling_maps}). To compile a circuit for such a coupling graph using \texttt{staq}, a \texttt{staq} device object with the qubit connectivity of the augmented coupling graph was created. The fidelity of each edge in the coupling graph was specified in this object, and the augmented edges were given poorer fidelities than the standard device edges. Initially chosen through trial and error, the standard device edges were assigned a fidelity of $0.9$ and the augmented edges were given a fidelity of $0.8$. The compiler converts these fidelities into weights to account for them in a graph route optimization algorithm, that can find the paths between specified nodes in the coupling graph whose edge weights sum to the lowest possible value. The fidelities are converted into weights by subtracting them from $1$, so the resulting weights for the standard and augmented edges were $0.1$ and $0.2$ respectively. This captures the fact that a remote CX gate uses two standard CX gates. If using an augmented edge was deemed optimal by the compiler, despite it's poorer fidelity, it would insert a standard CX gate in the circuit across this edge. This allowed us to determine where the remote CX gates should be placed. Once the circuit was fully compiled by \texttt{staq} for this device, a post processing step was then performed on the compiled circuit to determine where standard CX gates where inserted across the augmented edges. These standard CX gates were then replaced with remote CX gates in the compiled circuit. The final output is a circuit that has been compiled using both remote and standard CX gates.
    
    \section{Results} \label{sec:results}
        \subsection{Benchmarking Methodology}
            The method described in Sec.~\ref{sec:method} was implemented using Python. \texttt{staq}'s Steiner Tree mapper, with optimization level $2$, was used for compilation, and the post-processing and circuit manipulation was done with IBM's SDK QISkit~\cite{qiskit}. With this, circuits compiled using standard and remote CX gates were obtained for a set of benchmark circuits~\cite{amy_thesis, feynman, qedc_paper, qedc_repo}. For comparison, a second compiled circuit that used only standard gates, for a grid of the same size as that used for the remote CX gate compilation, was obtained for each benchmark. For this baseline, \texttt{staq} was used with a coupling graph that did not have any ancilla qubits reserved to form EPR pairs or any augmented edges added to it, and the optimization level was the same as that used for the remote CX gate compilation. Each compilation only needed to be performed once per circuit. \texttt{staq} is a deterministic compiler, and it was therefore not necessary to collect results over multiple trials.
            
    \subsection{Benchmark Results}
        \begin{figure}[]
            \centering
            \includegraphics[width=1\columnwidth]{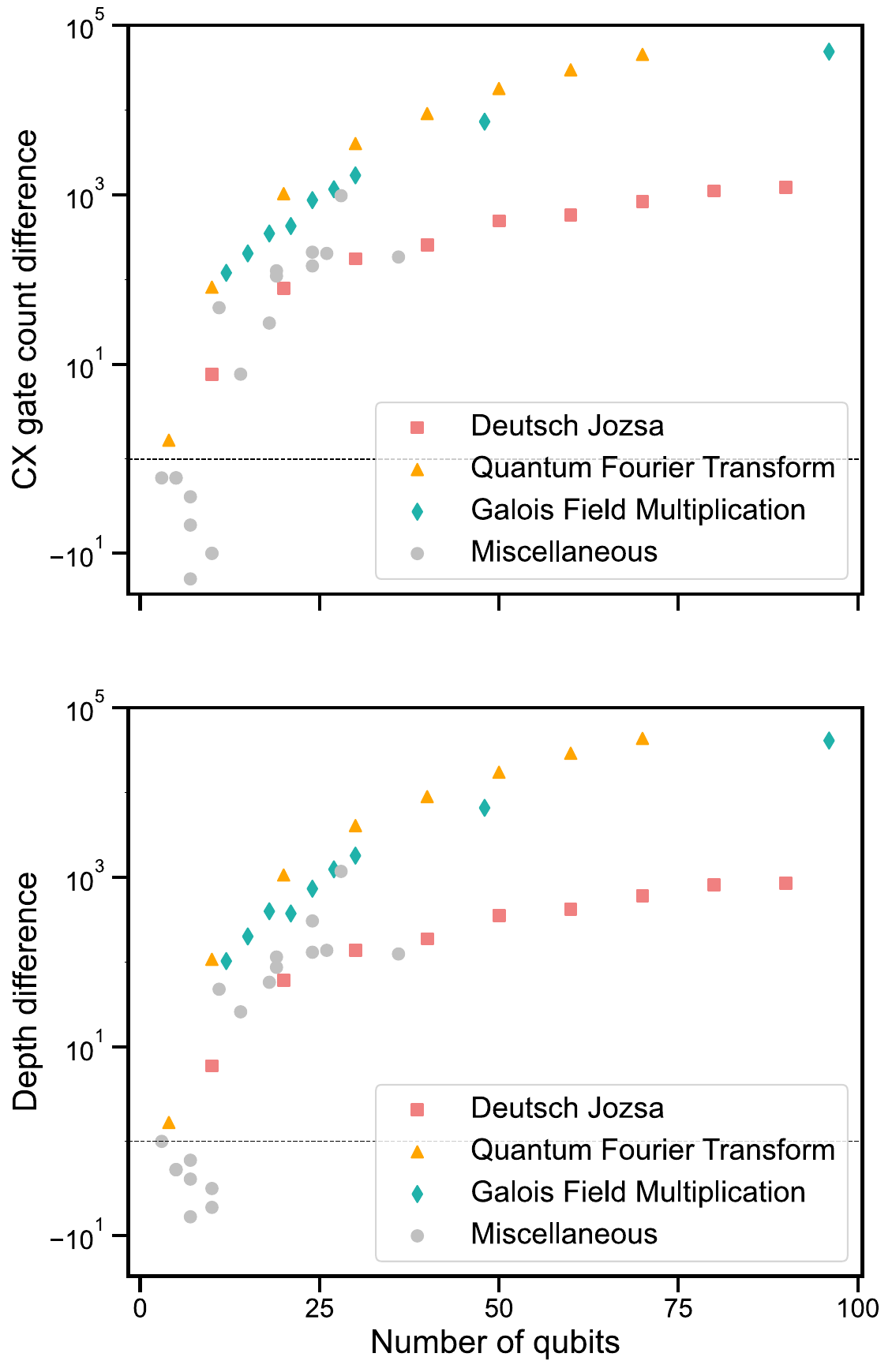}
            \caption{%
                Benchmark results plotted as the difference between the result obtained for the standard gate compilation and the remote gate compilation, as a function of the number of qubits of the circuit. A positive value indicates that the standard gate compilation received a higher value. 
                Note that the y-axis is logarithmically scaled. The results for some familiar benchmark circuits are identified. 
            }%
            \label{fig:plots}
        \end{figure}
        \begin{table}[]
                \centering
                \begin{tabular}{|l|ll|ll|}\hline
                    & \multicolumn{2}{c|}{Remote gate} & \multicolumn{2}{c|}{Standard}\\
                    \multicolumn{1}{|c|}{Circuit family}&\multicolumn{1}{c}{\#CX gates}&\multicolumn{1}{c|}{Depth}&\multicolumn{1}{c}{\#CX gates}&\multicolumn{1}{c|}{Depth}\\
                    \hline\hline
                    \begin{filecontents*}{data/stat_summary.csv}
a,b,c,d,e
Deutsch Jozsa,+505.5\%,+329.9\%,+1372.3\%,+922.7\%
Quantum FT,+567.1\%,+4057.6\%,+1551.1\%,+15863.9\%
GF Multiplier,+534.7\%,+854.5\%,+816.1\%,+1639.9\%
Miscellaneous,+201.8\%,+140.9\%,+226.3\%,+175.6\%
\end{filecontents*}
\csvreader[head to column names, late after line = \\]{data/stat_summary.csv}{}{\a & \b & \c &\d&\e}
                    \hline
                \end{tabular}
                \caption{%
                    Summary of the average overheads for the remote gate and standard compilation schemes. The percentage increase in gate counts and depths of the compiled circuits are calculated with respect to the values of the uncompiled circuits. Averages are taken within some familiar circuit families.
                }%
                \label{tab:stat_sum}
            \end{table}
        
        For each metric of interest, Figure~\ref{fig:plots} plots the difference between the result obtained for the standard gate compilation and the remote CX gate compilation for each circuit. A positive value indicates that the remote CX gate compilation achieved a lower value, and therefore improved on the result obtained using standard gates only. This improvement is observed for circuits with $10$ qubits or more, in both the standard CX gate counts and depths of the compiled circuits. The improvement scales with the size of the circuits, and in some cases reduces overheads by several orders of magnitude. The results for some familiar benchmark circuits are identified, and there is a clear separation between how different families of circuits performed under this model. 

    \section{Discussion}
        These results demonstrate that using remote CX gates for qubit routing can be highly advantageous, as they can significantly reduce both the standard CX gate counts and depths of the compiled circuits. The lower limit on the circuit size for which an advantage can be observed when remote gates are used is only $10$ qubits, which allows us to anticipate that most circuits for which there are practical use cases  would likely benefit from the use of remote CX gates. Despite the fact that the number of EPR pairs used scales linearly as the number of data qubits scales quadratically, the improvement imparted by the use of remote gates grows with the size of the circuits. This demonstrates that the use of remote gates for qubit routing can become more effective as architectures grow in size.

        There is a noticeable difference in the performance of circuits from different families, particularly in how the advantage scales as the circuits grow in size. The most significant advantage is observed for quantum Fourier transform (QFT) circuits. The physical qubit connectivity that would be necessary to satisfy all the two qubit gates between virtual qubits in a QFT circuit is that of a complete graph. In a complete graph, every node is connected by an edge to every other node in the graph. As QFT circuits grow in size, their virtual qubit connectivity differs more and more drastically from that of a square grid. As shown in Table~\ref{tab:stat_sum}, the gate count and depth overheads incurred when routing QFT circuits for this topology is very high. It is consistent that access to long-range connectivity would alleviate this overhead more significantly for QFT circuits than it would for Galois-Field (GF) multiplier and Deutsch-Jozsa (DJ) circuits, which have a lower virtual qubit connectivity. Unlike QFT circuits, the connectivity of GF multiplier circuits does not correspond to a characterized graph type. Nonetheless, it can be described as a graph where each node is connected to roughly $\frac{2}{3}$ of the other nodes in the graph, which imparts a lower connectivity than that of a QFT circuit. DJ circuits have the connectivity of a wheel graph, where there is one node that is connected to all the other nodes in the graph. The remaining nodes are each only connected to this one node, and these circuits therefore have a much lower virtual qubit connectivity than both the QFT and GF multiplier circuits. These characteristics likely factor into the rate at which the advantage of using remote CX gates, and having access to long-range connectivity, scales for each circuit type.

        In our analysis, we only considered using the distributed EPR pairs to perform remote CX gates, which are a subset of the possible operations that can use resource states and measurements to augment the connectivity of physical qubits. Teleportation has proven effective in this effort, particularly when combined with remote CX gates~\cite{tham2022}. However, integrating teleportation with an existing compiler presented a number of challenges. As an example, consider the case where two qubits in the device need to perform a two qubit gate, and can be brought closer to one another using teleportation. Whether it would be more advantageous to teleport one towards the other, or vice-versa, would need to be decided. Once this is decided, it is not obvious how to manage the teleported qubit after the gate is performed. The qubit can be teleported back to it's original position, which increases the incurred overhead of using teleportation. It can also remain at its new position, and the mapping of physical to virtual qubits can be updated. However, since a resource state generating device is being used to entangle qubits, the ancilla qubit which now contains the teleported qubit would have to be `freed up' so that the pair can be re-entangled. These decisions, among many others, would have had to be optimized within the constraints imposed by existing qubit routing techniques.
        
        Many of the intricacies of building an optimizing compiler from scratch were circumvented by using an existing compiler. Since the performance of the model was measured in terms of the resulting standard CX gate counts and depths of the compiled circuits, it was crucial that the compiler account for the increased cost of using a remote gate. This was achieved through the fidelity weighted routing feature of \texttt{staq}. Choosing the correct weights for the augmented coupling graph ensured that the trade-off between the increased cost of using a remote gate and the advantage of being able to perform the gate on distant qubits was properly balanced.
        
        While the introduction of the augmented edges in the coupling graph proved to be a useful way to account for the possibility of performing a remote gate, certain augmented edges use the same EPR pairs. These pairs can only be used for one remote gate at a time, however the compiler may have inserted standard CX gates in parallel across augmented edges that use the same EPR pair. With our implementation, these CX gates would have been performed in series. This may have lead to larger depths than anticipated by the compiler, however this did not appear to affect the results. 
        On average, $6\%$ of the CX operations in the compiled circuits were remote, and it is unlikely that it often tried to perform multiple remote gates simultaneously. Nonetheless, the method we described here provides a tool to further investigate the advantages and trade-offs of using EPR mediated operations for qubit routing.

    \section{Conclusion}
        We have developed an approach that leverages an existing state-of-the-art, heuristic based compiler, to optimize the use of both standard gates and EPR mediated remote CX gates to route circuits. This method allows us to explore the potential benefits and trade-offs of employing EPR-mediated operations for qubit routing. Our results show that by employing remote CX gates, we can significantly reduce the standard CX gate counts and depths of compiled benchmark circuits. The scale of this advantage is circuit dependent, and grows with the size of the architecture. These results highlight the potential value in creating compilation tools that can incorporate EPR mediated operations efficiently. Additionally, it prompts us to consider how to formally characterize and understand the discrepancy between the results obtained for different circuit families.

   \newpage
    \bibliographystyle{IEEEtran}
    \bibliography{biblio} 
\newpage
    \appendix
    \begin{table}[H] \onecolumn
        \centering
        \begin{tabular}{|l|rrr|rrr|rr|}\hline
            &\multicolumn{3}{c|}{Original}&\multicolumn{3}{c|}{ Remote gate compilation} &\multicolumn{2}{c|}{Standard compilation}\\ Circuit & \# qubits & \# CX& Depth& \# CX& Depth & \# rem. CX& \# CX& Depth\\\hline\hline
            \begin{filecontents*}{data/final_data.csv}
a,b,c,d,e,f,g,h,i
adder,10,65,100,103,98,11,93,91
adder\_8,24,409,252,974,553,128,1186,859
barenco\_tof\_10,19,192,339,310,288,9,438,403
barenco\_tof\_3,5,24,45,37,37,0,35,34
barenco\_tof\_4,7,48,87,75,74,0,68,66
bigadder,18,130,153,253,199,50,284,257
dj\_10,10,9,12,24,20,0,33,28
dj\_20,20,19,22,61,46,1,141,107
dj\_30,30,29,32,114,77,4,291,215
dj\_40,40,39,42,211,158,2,469,346
dj\_50,50,49,52,234,158,24,729,512
dj\_60,60,59,62,380,254,57,957,672
dj\_70,70,69,72,394,276,8,1235,883
dj\_80,80,79,82,458,280,44,1575,1096
dj\_90,90,89,92,672,495,14,1907,1341
gf$2^{10}$\_mult,30,609,308,3376,2109,234,5080,3909
gf$2^{16}$\_mult,48,1581,522,10362,6427,559,17673,13009
gf$2^{32}$\_mult,96,6268,1066,56310,32628,3881,105207,73470
gf$2^{4}$\_mult,12,99,108,340,280,14,461,383
gf$2^{5}$\_mult,15,154,138,542,406,23,747,607
gf$2^{6}$\_mult,18,221,176,1038,738,36,1390,1136
gf$2^{7}$\_mult,21,300,210,1554,1160,43,1985,1535
gf$2^{8}$\_mult,24,405,250,1973,1399,71,2843,2133
gf$2^{9}$\_mult,27,494,274,2638,1723,157,3808,2964
hwb6,7,116,169,329,285,12,309,283
mod\_adder\_1024,28,1702,2668,4817,3995,625,5798,5169
mod\_red\_21,11,105,162,211,183,4,258,231
mod5\_4,5,28,49,42,42,0,40,39
qcla\_adder\_10,36,215,78,637,336,98,823,461
qcla\_com\_7,24,174,88,400,194,30,546,325
qcla\_mod\_7,26,366,219,924,545,43,1129,683
qft,4,12,23,18,18,0,20,20
qft\_10,10,45,19,168,113,7,250,221
qft\_20,20,190,39,859,594,33,1897,1658
qft\_30,30,435,59,2392,1537,182,6428,5605
qft\_40,40,780,79,5218,3512,362,14322,12424
qft\_50,50,1225,99,8342,5460,527,26314,22761
qft\_60,60,1770,119,13777,8769,1376,43600,37731
qft\_70,70,2415,139,21328,14359,1579,66730,57665
rc\_adder\_6,14,91,108,178,118,17,187,144
tof\_10,19,102,180,188,178,0,299,265
tof\_3,5,18,33,32,32,0,30,29
tof\_4,7,30,54,48,46,0,44,42
vbe\_adder\_3,10,62,82,91,73,2,81,68
W-state,3,9,21,19,17,2,17,17
\end{filecontents*}
\csvreader[head to column names, late after line = \\]{data/final_data.csv}{}{%
                \a & \b & \c & \d & \e & \f & \g & \h & \i
            }%
            \hline
        \end{tabular}
        \caption{%
            A summary of the benchmark results obtained. Listed are the CX gate counts and depths for the original uncompiled benchmark circuits and for both the standard gate and remote CX gate compilations, as well as the number of remote gates used for the remote CX gate compilation
        }%
        \label{tab:final_data}
    \end{table}

\end{document}

%% file: figs/remote_cx.tex
\begin{quantikz}
\lstick{$q_{0}$}&\ctrl{3}&\qw\\
\\
\\
\lstick{$q_{1}$}&\targ{}&\qw
\end{quantikz}=\begin{quantikz}
\lstick{$q_{0}$} & \ctrl{1}&\qw&\qw&\gate{Z}&\qw \\
\lstick[wires = 2]{$\ket{\phi^{+}}$} & \targ{}& \qw&\meter{} \vcw{2}\\
& \ctrl{1}&\gate{H}&\qw&\meter{} \vcw{-2} \\
\lstick{$q_{1}$}&\targ{}&\qw&\gate{X}&\qw&\qw 
\end{quantikz}

%% file: figs/tikz_hole.tex
\begin{tikzpicture}
[scale=1.2,auto=left,every node/.style={circle,fill=white, draw=black, minimum size = 0.8 cm}]
  \node (n0) at (0, 2) {$q_{0}$};
  \node (n1) at (1, 2) {$q_{1}$};
  \node (n2) at (2, 2) {$q_{2}$};
  
  \node (n3) at (0, 1) {$q_{3}$};
  \node[fill=black, text=white] (n4) at (1, 1) {$q_{4}$};
  \node (n5) at (2, 1) {$q_{5}$};
  
  \node (n6) at (0, 0) {$q_{6}$};
  \node (n7) at (1, 0) {$q_{7}$};
  \node (n8) at (2, 0) {$q_{8}$};

  \foreach \from/\to in {%
    n0/n1, n1/n2,%
    n3/n4, n4/n5,%
    n0/n3, n3/n6,%
    n1/n4, n4/n7,%
    n2/n5%
    }
    \draw (\from) -- (\to);
    \foreach \from/\to in {%
    n8/n5, n6/n7, n7/n8, n3/n6%
    }
    \draw[->, very thick] (\from) -- (\to);
\end{tikzpicture}

%% file: figs/tikz_no_hole.tex
\begin{tikzpicture}[scale=1.2,auto=left,every node/.style={circle,fill=white, draw=black, minimum size = 0.8 cm}]
  \node (n0) at (0, 2) {$q_{0}$};
  \node (n1) at (1, 2) {$q_{1}$};
  \node (n2) at (2, 2) {$q_{2}$};
  
  \node (n3) at (0, 1) {$q_{3}$};
  \node (n4) at (1, 1) {$q_{4}$};
  \node (n5) at (2, 1) {$q_{5}$};
  
  \node (n6) at (0, 0) {$q_{6}$};
  \node (n7) at (1, 0) {$q_{7}$};
  \node (n8) at (2, 0) {$q_{8}$};

  \foreach \from/\to in {%
    n0/n1, n1/n2,%
    n3/n4, n4/n5,%
    n6/n7, n7/n8,%
    n0/n3, n3/n6,%
    n1/n4, n4/n7,%
    n2/n5, n5/n8%
    }
    \draw (\from) -- (\to);
    \foreach \from/\to in {%
    n4/n5, n3/n4%
    }
    \draw [->, very thick] (\from) -- (\to);
\end{tikzpicture}

%% file: figs/tikz_epr.tex
\begin{tikzpicture}
\begin{scope}[scale=1.2, auto=left, every node/.style={circle,fill=white, draw=black, minimum size = 0.8 cm}]
  \draw[dotted, very thick] (0, 4) arc [radius=3.5, start angle=125, end angle= 55];
    \draw[dotted, very thick] (0, 0) arc [radius=3.5, start angle=235, end angle= 305];
    \draw[dotted, very thick] (1, 4) arc [radius=4, start angle=150, end angle= 210];
    \draw[dotted, very thick] (3, 4) arc [radius=4, start angle=30, end angle= -30];
  \node[fill=black, text=white] (n0) at (0, 4) {$q_{0}$};
  \node[fill=black, text=white] (n1) at (1, 4) {$q_{1}$};
  \node (n2) at (2, 4) {$q_{2}$};
  \node[fill=black, text=white] (n3) at (3, 4) {$q_{3}$};
  \node[fill=black, text=white] (n4) at (4, 4) {$q_{4}$};

  \node (n5) at (0, 3) {$q_{5}$};
  \node (n6) at (1, 3) {$q_{6}$};
  \node (n7) at (2, 3) {$q_{7}$};
  \node (n8) at (3, 3) {$q_{8}$};
  \node (n9) at (4, 3) {$q_{9}$};
  
  \node (n10) at (0, 2) {$q_{10}$};
  \node (n11) at (1, 2) {$q_{11}$};
  \node (n12) at (2, 2) {$q_{12}$};
  \node (n13) at (3, 2) {$q_{13}$};
  \node (n14) at (4, 2) {$q_{14}$};

  \node (n15) at (0, 1) {$q_{15}$};
  \node (n16) at (1, 1) {$q_{16}$};
  \node (n17) at (2, 1) {$q_{17}$};
  \node (n18) at (3, 1) {$q_{18}$};
  \node (n19) at (4, 1) {$q_{19}$};
  
  \node (n20)[fill=black, text=white] at (0, 0) {$q_{20}$};
  \node (n21)[fill=black, text=white] at (1, 0) {$q_{21}$};
  \node (n22) at (2, 0) {$q_{22}$};
  \node (n23)[fill=black, text=white] at (3, 0) {$q_{23}$};
  \node (n24)[fill=black, text=white] at (4, 0) {$q_{24}$};
  \foreach \from/\to in {%
    n0/n1, n1/n2, n2/n3, n3/n4,%
    n5/n6, n6/n7, n7/n8, n8/n9,%
    n10/n11, n11/n12, n12/n13, n13/n14,%
    n15/n16, n16/n17, n17/n18, n18/n19,%
    n20/n21, n21/n22, n22/n23, n23/n24,%
    n0/n5, n5/n10, n10/n15, n15/n20,%
    n1/n6, n6/n11, n11/n16, n16/n21,%
    n2/n7, n7/n12, n12/n17, n17/n22,%
    n3/n8, n8/n13, n13/n18, n18/n23,%
    n4/n9, n9/n14, n14/n19, n19/n24%
    }
    \draw (\from) -- (\to);
    \draw[->] (4.5, 2)--(6.2, 2);
\end{scope}
\begin{scope}[xshift = 8 cm, scale=1.2, auto=left, every node/.style={circle,fill=white, draw=black, minimum size = 0.8 cm}]
    \draw[black, very thick] (0,3) arc [radius=4, start angle=120, end angle= 60];
    \draw[black, very thick] (1,3) arc [radius=1.4, start angle=135, end angle= 225];
    \draw[black, very thick] (2, 4) arc [radius=4, start angle=30, end angle= -30];
    \draw[black, very thick] (3, 3) arc [radius=1.4, start angle=45, end angle= -45];
    \draw[black, very thick] (0, 1) arc [radius=4, start angle=240, end angle= 300];
    \draw[black, very thick] (1, 3) arc [radius=6, start angle=30, end angle= 0];
    \draw[black, very thick] (1, 1) arc [radius=6, start angle=-30, end angle= 0];
    \draw[black, very thick] (3, 3) arc [radius=6, start angle=150, end angle= 180];
    \draw[black, very thick] (3, 1) arc [radius=6, start angle=210, end angle= 180];
  \node (n2) at (2, 4) {$q_{2}$};

  \node (n5) at (0, 3) {$q_{5}$};
  \node (n6) at (1, 3) {$q_{6}$};
  \node (n7) at (2, 3) {$q_{7}$};
  \node (n8) at (3, 3) {$q_{8}$};
  \node (n9) at (4, 3) {$q_{9}$};
  
  \node (n10) at (0, 2) {$q_{10}$};
  \node (n11) at (1, 2) {$q_{11}$};
  \node (n12) at (2, 2) {$q_{12}$};
  \node (n13) at (3, 2) {$q_{13}$};
  \node (n14) at (4, 2) {$q_{14}$};

  \node (n15) at (0, 1) {$q_{15}$};
  \node (n16) at (1, 1) {$q_{16}$};
  \node (n17) at (2, 1) {$q_{17}$};
  \node (n18) at (3, 1) {$q_{18}$};
  \node (n19) at (4, 1) {$q_{19}$};
  
  \node (n22) at (2, 0) {$q_{22}$};
  
  \foreach \from/\to in {%
    n5/n6, n6/n7, n7/n8, n8/n9,%
    n10/n11, n11/n12, n12/n13, n13/n14,%
    n15/n16, n16/n17, n17/n18, n18/n19,%
    n5/n10, n10/n15,%
    n6/n11, n11/n16,%
    n2/n7, n7/n12, n12/n17, n17/n22,%
    n8/n13, n13/n18,%
    n9/n14, n14/n19%
    }
    \draw (\from) -- (\to);
\end{scope}
\end{tikzpicture}

%% file: main.bbl
\begin{thebibliography}{10}
\providecommand{\url}[1]{#1}
\csname url@samestyle\endcsname
\providecommand{\newblock}{\relax}
\providecommand{\bibinfo}[2]{#2}
\providecommand{\BIBentrySTDinterwordspacing}{\spaceskip=0pt\relax}
\providecommand{\BIBentryALTinterwordstretchfactor}{4}
\providecommand{\BIBentryALTinterwordspacing}{\spaceskip=\fontdimen2\font plus
\BIBentryALTinterwordstretchfactor\fontdimen3\font minus
  \fontdimen4\font\relax}
\providecommand{\BIBforeignlanguage}[2]{{%
\expandafter\ifx\csname l@#1\endcsname\relax
\typeout{** WARNING: IEEEtran.bst: No hyphenation pattern has been}%
\typeout{** loaded for the language `#1'. Using the pattern for}%
\typeout{** the default language instead.}%
\else
\language=\csname l@#1\endcsname
\fi
#2}}
\providecommand{\BIBdecl}{\relax}
\BIBdecl

\bibitem{preskill_2018}
\BIBentryALTinterwordspacing
J.~Preskill, ``Quantum computing in the {NISQ} era and beyond,''
  \emph{Quantum}, vol.~2, p.~79, 2018. [Online]. Available:
  \url{https://doi.org/10.22331%2Fq-2018-08-06-79}
\BIBentrySTDinterwordspacing

\bibitem{siraichi_2018}
\BIBentryALTinterwordspacing
M.~Y. Siraichi, V.~F.~d. Santos, C.~Collange, and F.~M.~Q. Pereira, ``Qubit
  allocation,'' in \emph{Proceedings of the 2018 International Symposium on
  Code Generation and Optimization}, ser. CGO 2018.\hskip 1em plus 0.5em minus
  0.4em\relax New York, NY, USA: Association for Computing Machinery, 2018, p.
  113–125. [Online]. Available: \url{https://doi.org/10.1145/3168822}
\BIBentrySTDinterwordspacing

\bibitem{lau_2022}
J.~W. Lau, K.~H. Lim, H.~Shrotriya, and L.~C. Kwek, ``Nisq computing: Where are
  we and where do we go?'' \emph{AAPPS Bulletin}, vol.~32, no.~1, 2022.

\bibitem{sabre}
\BIBentryALTinterwordspacing
G.~Li, Y.~Ding, and Y.~Xie, ``Tackling the qubit mapping problem for nisq-era
  quantum devices,'' in \emph{Proceedings of the Twenty-Fourth International
  Conference on Architectural Support for Programming Languages and Operating
  Systems}, ser. ASPLOS '19.\hskip 1em plus 0.5em minus 0.4em\relax New York,
  NY, USA: Association for Computing Machinery, 2019, p. 1001–1014. [Online].
  Available: \url{https://doi.org/10.1145/3297858.3304023}
\BIBentrySTDinterwordspacing

\bibitem{cnot_synth}
A.~Kissinger and A.~M. van~de Griend, ``Cnot circuit extraction for
  topologically-constrained quantum memories,'' 2019.

\bibitem{tket}
\BIBentryALTinterwordspacing
A.~Cowtan, S.~Dilkes, R.~Duncan, A.~Krajenbrink, W.~Simmons, and S.~Sivarajah,
  ``{On the Qubit Routing Problem},'' in \emph{14th Conference on the Theory of
  Quantum Computation, Communication and Cryptography (TQC 2019)}, ser. Leibniz
  International Proceedings in Informatics (LIPIcs), W.~van Dam and
  L.~Mancinska, Eds., vol. 135.\hskip 1em plus 0.5em minus 0.4em\relax
  Dagstuhl, Germany: Schloss Dagstuhl--Leibniz-Zentrum fuer Informatik, 2019,
  pp. 5:1--5:32. [Online]. Available:
  \url{http://drops.dagstuhl.de/opus/volltexte/2019/10397}
\BIBentrySTDinterwordspacing

\bibitem{zulehner2018}
A.~Zulehner, A.~Paler, and R.~Wille, ``An efficient methodology for mapping
  quantum circuits to the ibm qx architectures,'' \emph{IEEE Transactions on
  Computer-Aided Design of Integrated Circuits and Systems}, vol.~38, no.~7,
  pp. 1226--1236, 2018.

\bibitem{pyzx}
A.~Kissinger and J.~van~de Wetering, ``{PyZX: Large Scale Automated
  Diagrammatic Reasoning},'' in \emph{{\rm Proceedings 16th International
  Conference on} Quantum Physics and Logic, {\rm Chapman University, Orange,
  CA, USA., 10-14 June 2019}}, ser. Electronic Proceedings in Theoretical
  Computer Science, B.~Coecke and M.~Leifer, Eds., vol. 318.\hskip 1em plus
  0.5em minus 0.4em\relax Open Publishing Association, 2020, pp. 229--241.

\bibitem{maslov}
D.~Maslov, ``Basic circuit compilation techniques for an ion-trap quantum
  machine,'' \emph{New Journal of Physics}, vol.~19, no.~2, p. 023035, 2017.

\bibitem{gheorghiu}
V.~Gheorghiu, J.~Huang, S.~M. Li, M.~Mosca, and P.~Mukhopadhyay, ``Reducing the
  cnot count for clifford+t circuits on nisq architectures,'' \emph{IEEE
  Transactions on Computer-Aided Design of Integrated Circuits and Systems},
  vol.~42, no.~6, pp. 1873--1884, 2023.

\bibitem{leymann2020}
\BIBentryALTinterwordspacing
F.~Leymann and J.~Barzen, ``The bitter truth about gate-based quantum
  algorithms in the nisq era,'' \emph{Quantum Science and Technology}, vol.~5,
  no.~4, p. 044007, 2020. [Online]. Available:
  \url{https://app.dimensions.ai/details/publication/pub.1130078978}
\BIBentrySTDinterwordspacing

\bibitem{herbert2020}
S.~Herbert, ``On the depth overhead incurred when running quantum algorithms on
  near-term quantum computers with limited qubit connectivity,'' \emph{Quantum
  Information and Computation}, vol.~20, no. 9 \& 10, p. 787–806, 2020.

\bibitem{teleportation}
\BIBentryALTinterwordspacing
C.~H. Bennett, G.~Brassard, C.~Cr\'epeau, R.~Jozsa, A.~Peres, and W.~K.
  Wootters, ``Teleporting an unknown quantum state via dual classical and
  einstein-podolsky-rosen channels,'' \emph{Phys. Rev. Lett.}, vol.~70, pp.
  1895--1899, 1993. [Online]. Available:
  \url{https://link.aps.org/doi/10.1103/PhysRevLett.70.1895}
\BIBentrySTDinterwordspacing

\bibitem{collins}
\BIBentryALTinterwordspacing
D.~Collins, N.~Linden, and S.~Popescu, ``Nonlocal content of quantum
  operations,'' \emph{Phys. Rev. A}, vol.~64, p. 032302, 2001. [Online].
  Available: \url{https://link.aps.org/doi/10.1103/PhysRevA.64.032302}
\BIBentrySTDinterwordspacing

\bibitem{nielson_r_cx}
\BIBentryALTinterwordspacing
M.~A. Nielsen and I.~L. Chuang, ``Programmable quantum gate arrays,''
  \emph{Phys. Rev. Lett.}, vol.~79, pp. 321--324, 1997. [Online]. Available:
  \url{https://link.aps.org/doi/10.1103/PhysRevLett.79.321}
\BIBentrySTDinterwordspacing

\bibitem{sorensen}
\BIBentryALTinterwordspacing
A.~S\o{}rensen and K.~M\o{}lmer, ``Error-free quantum communication through
  noisy channels,'' \emph{Phys. Rev. A}, vol.~58, pp. 2745--2749, 1998.
  [Online]. Available: \url{https://link.aps.org/doi/10.1103/PhysRevA.58.2745}
\BIBentrySTDinterwordspacing

\bibitem{alon1993routing}
\BIBentryALTinterwordspacing
N.~Alon, F.~R. Chung, and R.~L. Graham, ``Routing permutations on graphs via
  matchings,'' in \emph{Proceedings of the Twenty-Fifth Annual ACM Symposium on
  Theory of Computing}, ser. STOC '93.\hskip 1em plus 0.5em minus 0.4em\relax
  New York, NY, USA: Association for Computing Machinery, 1993, p. 583–591.
  [Online]. Available: \url{https://doi.org/10.1145/167088.167239}
\BIBentrySTDinterwordspacing

\bibitem{devulapalli2022}
D.~Devulapalli, E.~Schoute, A.~Bapat, A.~M. Childs, and A.~V. Gorshkov,
  ``Quantum routing with teleportation,'' 2022.

\bibitem{hillmich2021}
\BIBentryALTinterwordspacing
S.~Hillmich, A.~Zulehner, and R.~Wille, ``Exploiting quantum teleportation in
  quantum circuit mapping,'' in \emph{Proceedings of the 26th Asia and South
  Pacific Design Automation Conference}, ser. ASPDAC '21.\hskip 1em plus 0.5em
  minus 0.4em\relax New York, NY, USA: Association for Computing Machinery,
  2021, p. 792–797. [Online]. Available:
  \url{https://doi.org/10.1145/3394885.3431604}
\BIBentrySTDinterwordspacing

\bibitem{stephenson}
\BIBentryALTinterwordspacing
L.~J. Stephenson, D.~P. Nadlinger, B.~C. Nichol, S.~An, P.~Drmota, and T.~G.
  Ballance~\MakeLowercase{\textit{et al.}}, ``High-rate, high-fidelity
  entanglement of qubits across an elementary quantum network,'' \emph{Phys.
  Rev. Lett.}, vol. 124, p. 110501, 2020. [Online]. Available:
  \url{https://link.aps.org/doi/10.1103/PhysRevLett.124.110501}
\BIBentrySTDinterwordspacing

\bibitem{monroe}
\BIBentryALTinterwordspacing
C.~Monroe, R.~Raussendorf, A.~Ruthven, K.~R. Brown, P.~Maunz, L.-M. Duan, and
  J.~Kim, ``Large-scale modular quantum-computer architecture with atomic
  memory and photonic interconnects,'' \emph{Phys. Rev. A}, vol.~89, p. 022317,
  2014. [Online]. Available:
  \url{https://link.aps.org/doi/10.1103/PhysRevA.89.022317}
\BIBentrySTDinterwordspacing

\bibitem{remote_cx}
\BIBentryALTinterwordspacing
L.~Duan, B.~B. Blinov, D.~L. Moehring, and C.~R. Monroe, ``Scalable trapped ion
  quantum computation with a probabilistic ion-photon mapping,'' \emph{Quantum
  Information and Computation}, vol.~4, no.~3, pp. 165--173, 2004. [Online].
  Available: \url{https://doi.org/10.26421/QIC4.3-1}
\BIBentrySTDinterwordspacing

\bibitem{tham2022}
E.~Tham, I.~Khait, and A.~Brodutch, ``Quantum circuit optimization for multiple
  qpus using local structure,'' in \emph{2022 IEEE International Conference on
  Quantum Computing and Engineering (QCE)}.\hskip 1em plus 0.5em minus
  0.4em\relax New York, NY, USA: Institute of Electrical and Electronics
  Engineers, 2022, pp. 476--483.

\bibitem{botea_2018}
\BIBentryALTinterwordspacing
A.~Botea, A.~Kishimoto, and R.~Marinescu, ``On the complexity of quantum
  circuit compilation,'' in \emph{Proceedings of the International Symposium on
  Combinatorial Search}.\hskip 1em plus 0.5em minus 0.4em\relax Vienna,
  Austria: Association for the Advancement of Artificial Intelligence, 2018,
  pp. 138--142. [Online]. Available:
  \url{https://doi.org/10.1609/socs.v9i1.18463}
\BIBentrySTDinterwordspacing

\bibitem{ito2023}
T.~Ito, N.~Kakimura, N.~Kamiyama, Y.~Kobayashi, and Y.~Okamoto, ``Algorithmic
  theory of qubit routing,'' 2023.

\bibitem{nielsen_chuang_2010}
M.~A. Nielsen and I.~L. Chuang, \emph{Quantum Computation and Quantum
  Information: 10th Anniversary Edition}.\hskip 1em plus 0.5em minus
  0.4em\relax Cambridge University Press, 2010.

\bibitem{chow2021ibm}
J.~Chow, O.~Dial, and J.~Gambetta. (2021) Ibm quantum breaks the 100-qubit
  processor barrier.

\bibitem{arute2019quantum}
F.~Arute, K.~Arya, R.~Babbush, D.~Bacon, J.~C. Bardin, R.~Barends, R.~Biswas,
  S.~Boixo, F.~G. Brandao, D.~A. Buell \emph{et~al.}, ``Quantum supremacy using
  a programmable superconducting processor,'' \emph{Nature}, vol. 574, no.
  7779, pp. 505--510, 2019.

\bibitem{staq}
M.~Amy and V.~Gheorghiu, ``staq—a full-stack quantum processing toolkit,''
  \emph{Quantum Science and Technology}, vol.~5, no.~3, p. 034016, 2020.

\bibitem{Awschalom_2021}
\BIBentryALTinterwordspacing
D.~Awschalom, K.~K. Berggren, H.~Bernien, S.~Bhave, L.~D. Carr, and
  P.~Davids~\MakeLowercase{\textit{et al.}}, ``Development of quantum
  interconnects (quics) for next-generation information technologies,''
  \emph{PRX Quantum}, vol.~2, no.~1, Feb. 2021. [Online]. Available:
  \url{http://dx.doi.org/10.1103/PRXQuantum.2.017002}
\BIBentrySTDinterwordspacing

\bibitem{Hucul_2014}
\BIBentryALTinterwordspacing
D.~Hucul, I.~V. Inlek, G.~Vittorini, C.~Crocker, S.~Debnath, S.~M. Clark, and
  C.~Monroe, ``Modular entanglement of atomic qubits using photons and
  phonons,'' \emph{Nature Physics}, vol.~11, no.~1, p. 37–42, Nov. 2014.
  [Online]. Available: \url{http://dx.doi.org/10.1038/nphys3150}
\BIBentrySTDinterwordspacing

\bibitem{transduc}
\BIBentryALTinterwordspacing
C.~Wang, I.~Gonin, A.~Grassellino, S.~Kazakov, A.~Romanenko, V.~P. Yakovlev,
  and S.~Zorzetti, ``High-efficiency microwave-optical quantum transduction
  based on a cavity electro-optic superconducting system with long coherence
  time,'' \emph{npj Quantum Information}, vol.~8, no.~1, p. 149, 2022.
  [Online]. Available: \url{https://doi.org/10.1038/s41534-022-00664-7}
\BIBentrySTDinterwordspacing

\bibitem{qiskit}
{Qiskit contributors}, ``Qiskit: An open-source framework for quantum
  computing,'' 2023.

\bibitem{amy_thesis}
\BIBentryALTinterwordspacing
M.~Amy, ``Formal methods in quantum circuit design,'' Ph.D. dissertation,
  University of Waterloo, Waterloo, Ontario, Canada, 2019. [Online]. Available:
  \url{http://hdl.handle.net/10012/14480}
\BIBentrySTDinterwordspacing

\bibitem{feynman}
------, ``Quantum circuit analysis toolkit,''
  \url{https://github.com/meamy/feynman}, accessed: July 5, 2022.

\bibitem{qedc_paper}
T.~Lubinski, S.~Johri, P.~Varosy, J.~Coleman, L.~Zhao, J.~Necaise, C.~H.
  Baldwin, K.~Mayer, and T.~Proctor, ``Application-oriented performance
  benchmarks for quantum computing,'' \emph{IEEE Transactions on Quantum
  Engineering}, vol.~4, pp. 1--32, 2023.

\bibitem{qedc_repo}
Q.~E. D. C. T. A.~C. on~Standards and P.~Metrics, ``Application-oriented
  performance benchmarks for quantum computing,''
  \url{https://github.com/SRI-International/QC-App-Oriented-Benchmarks},
  accessed: November 1, 2022.

\end{thebibliography}
